\documentclass[
reprint,superscriptaddress,
 amsmath,amssymb,
 aps,
]{revtex4-2}

\usepackage{amsmath}
\usepackage{bm}
\usepackage{graphicx}% Include figure files
\usepackage{dcolumn}% Align table columns on decimal point
\usepackage{bm}% bold math
\usepackage{hyperref}% add hypertext capabilities
\usepackage{epstopdf}

\begin{document}

\title{A new complex fluid flow phenomenon: Bubbles-on-a-String}

\author{Thomas P. John}
\affiliation{
Department of Chemical Engineering, The University of Manchester, UK
}
\author{Jack. R. C. King}
\email{jack.king@manchester.ac.uk}
\affiliation{
Department of Mechanical and Aerospace Engineering, The University of Manchester, UK
}
\author{Elliott Sutton}
\affiliation{
Department of Chemical Engineering, The University of Manchester, UK
}
\author{Steven J. Lind}
\email{linds1@cardiff.ac.uk}
\affiliation{School of Engineering, Cardiff University, UK}
\author{Cláudio P. Fonte}
\email{claudio.fonte@manchester.ac.uk}
\affiliation{
Department of Chemical Engineering, The University of Manchester, UK
}%
\date{\today}


\begin{abstract}
A liquid jet plunging into a quiescent bath of the same liquid is a fundamental fluid mechanical problem underpinning a range of processes in industry and the natural world. Significant attention has been given to the study of plunging laminar Newtonian jets and the associated air entrainment that can occur. However, there have been very few (if any) studies devoted to the equivalent case for non-Newtonian viscoelastic liquids. Here we consider the laminar plunging and associated air entrainemnt of a shear thinning viscoelastic jet into a still bath of the same liquid. We describe a previously unreported phenomenon, that we call ``bubbles-on-a-string'' (BUoaS), consisting of multiple stable toroidal bubbles rising co-axially around the submerged jet. In a qualitative sense, this new observation is akin to an inverse version of the well-known rheological phenomenon ``beads-on-a-string''. The BUoaS phenomenon is stable and repeatable and can be reproduced to a lesser extent in Newtonian surfactant solutions, indicating that low surface tension is key, but non-Newtonian rheology seems likely to provide the most favourable conditions for the onset of the phenomenon. A full characterisation and detailed study of this behaviour with accompanying numerical simulation is to follow in an upcoming publication.   
\end{abstract}

\maketitle

%\section{Introduction}
Liquid jets plunging into a still bath of the same liquid is a fundamental fluid mechanical process that underpins a range of related fluid entry problems across engineering, industry, and the natural world. Impinging liquid jet entry is important in industrial processes including printing, mixing~\cite{Tojo1982}, and filling. In oceanography, plunging jets are a critical part of the wave-breaking process, and form an important mechanism for air-sea gas exchange, and thereby influencing weather, climate, and ocean ecology~\cite{deike2022mass}. As the impinging jet enters the liquid bath, air is usually entrained with the amount of entrainment (in the case of water, for example), dependent on the Reynolds and Weber numbers. Generally, air entrainment can be desirable (e.g. plunging jet reactors) or undesirable (e.g. filling of containers), and the fundamental, ubiquitous nature of the process is such that it has been studied in detail in the Newtonian (mainly water) case for many years across both laminar and turbulent regimes. In the laminar regime of interest here, investigations have focused mainly on the degree of air entrainment~\cite{lin_1966} or the initial cusp instability responsible for the plunging~\cite{eggers_2001} and the lubricating air film~\cite{lorenceau_2004}.\\

\begin{figure}[ht]
\centering
\includegraphics[trim={0 0 10cm 0},clip,width=0.3\linewidth]{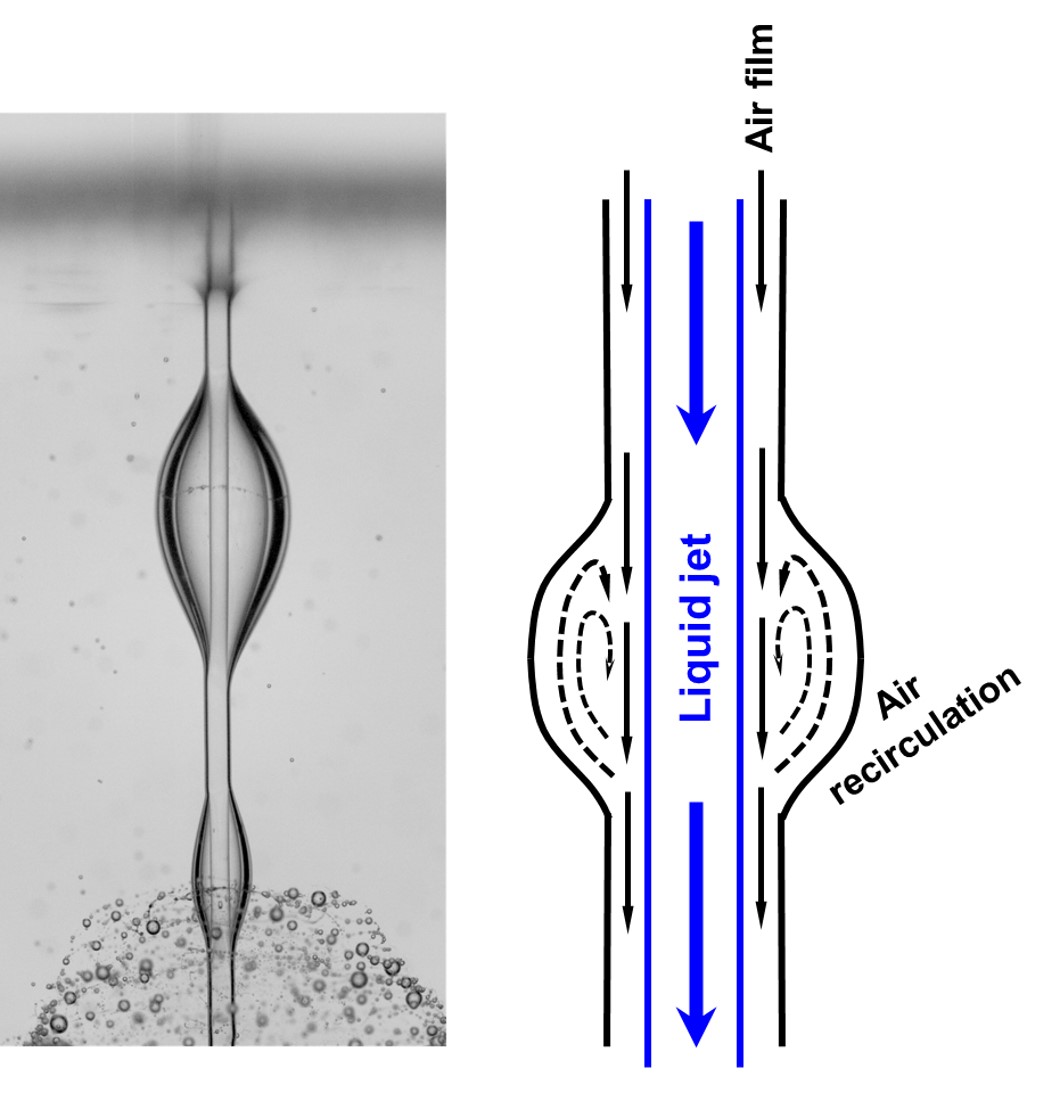}
%\hspace{1cm}
\includegraphics[width=0.45\linewidth]{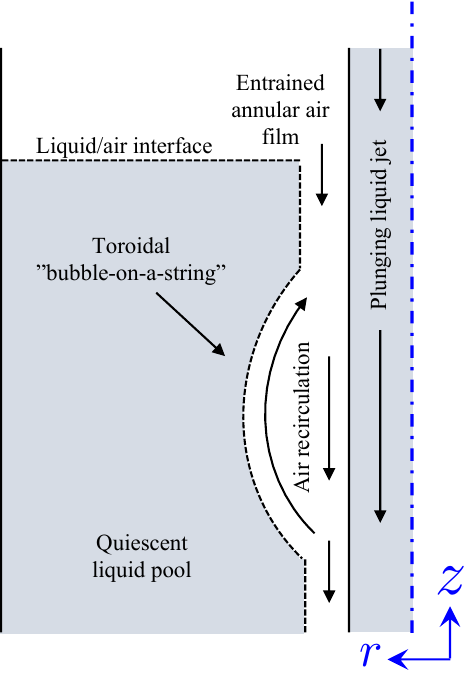}
\includegraphics[width=0.9\linewidth]{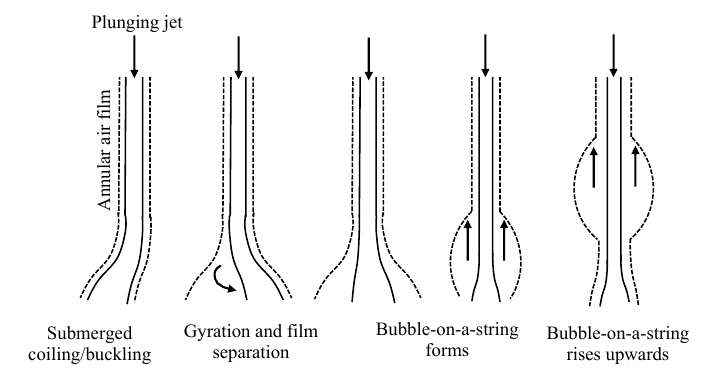}
\caption{Top Left: A close-up photograph of rising bubbles co-axial to a submerged liquid jet (bubbles-on-a-string). Top Right: A schematic (axisymmetric) of the expected flow directions within the air and liquid. Note that, in the left hand image, the black region at the edge of the bubble is a result of refraction, and the cloud of bubbles at the bottom of the image is caused by the initial plunging event and the shedding of the air layer at the bottom of the jet (which is below the lower boundary of the image). Bottom: Schematic timeline of the phenomenon.}
\label{fig:bubbles on string close up}
\end{figure}

In contrast, for the laminar non-Newtonian case, there is a dearth of studies in the literature, despite the prominence of non-Newtonian liquids across manufacturing and processing industries where jet impingement and plunging is a key process (e.g. in formulated consumer product mixing and container filling) and where entrained air content can impact product quality and cost. There have been many studies (computational and experimental) of viscoelastic jets generally, including on the Kaye effect~\cite{majmudar_2010,lee_2013,bonito_2016} and jet breakup~\cite{mcilroy_2013,harlen_2020}, but no dedicated studies to the authors' knowledge considering laminar jet plunging with air entrainment, although~\cite{lee_2013} sought to quantify air entrainment due to the Kaye effect. 

We address this gap in the literature by considering a shear-thinning viscoelastic laminar liquid jet plunging into a bath of the same liquid and studying experimentally the conditions under which air entrainment occurs. \textit{The principal outcome of this study is the identification of a previously unreported phenomenon consisting of multiple rising toroidal bubbles co-axial to the submerged jet (see Figure~\ref{fig:bubbles on string close up}). We term this phenomenon ``bubbles-on-a-string'' (BUoaS) due to its resemblance to the well-known ``beads-on-a-string'' phenomenon~\cite{Goldin_1969,clasen_2006,bhat_2010} observed in viscoelastic fluids.} 

%In \S~\ref{sec:results} we describe the bubbles-on-a-string phenomenon, the plunging dynamics, and the effect of changing fluid properties. Conclusions and items for further investigation follow in \S~\ref{sec:conc}. The experimental methods and materials used in this study are detailed in Appendix~\ref{sec:methods}. Movies showing the phenomenon are provided as ancillary files.

%\section{Results and Discussion}
%\label{sec:results}

%\subsection{The ``Bubbles-on-a-string'' phenomenon}\label{buoas}

\begin{figure*}%[ht]
\centering
\includegraphics[width=0.99\textwidth]{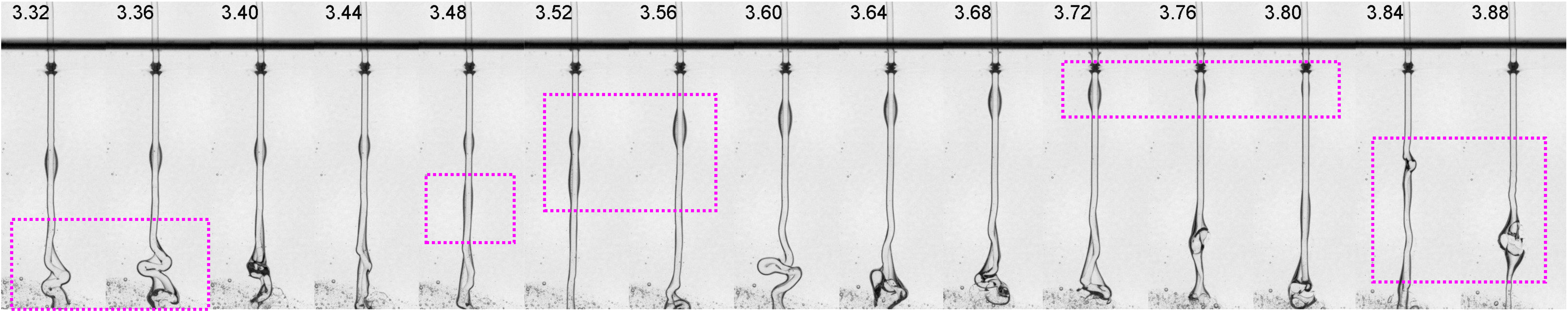}
\caption{The key sequence of events during the creation of ``bubbles-on-a-string''. Dashed boxes highlight, from left to right, 1 - gyration of the jet and resulting air film separation, 2 - bubble-on-a-string forms and travels upwards, 3 - bubble merges with existing bubble to form a single larger bubble, 4 - bubble collapses, 5 - bubble collapse disrupts the plunging.}
\label{fig:bos-sequence}
\end{figure*}

As the viscoelastic jet plunges into the liquid bath, a thin, effectively annular, air layer is entrained that encases the downward cylindrical jet. This laminar air entrainment has been shown (at least for Newtonian liquids) to be the result of a cusp instability which occurs at the meniscus where the falling jet impacts the quiescent pool \cite{lorenceau_2004}. As the vertical speed of the submerged jet slows with depth into the bath, conservation of mass demands the jet cross-sectional area expands, and the end of the jet is characterised by fanning out of the jet and the attached air layer, which breaks into smaller bubbles that are dispersed into the ambient liquid. In a similar manner to standard jet coiling/bulking, the end of the jet also undergoes a submerged coiling/buckling that perturbs, and can appear to separate from, the entrained air layer. This gyration of the jet and separation causes a local expansion of the air-layer volume, which in combination with local surface tension and fluid surface stresses results in the creation of a stable toroidal bubble in the air layer. This onset mechanism is shown schematically in the bottom image of Figure~\ref{fig:bubbles on string close up}. This bubble rises due to buoyancy along (and co-axial to) the column of the liquid jet, toward the liquid bath surface. As the end of the submerged jet continues to coil, this process is repeated periodically, creating multiple ``bubbles-on-a-string'' that exist for a sustained period of time. A close-up of such stark bubble behaviour is seen on the top-left of Figure \ref{fig:bubbles on string close up}. The top-right hand image shows a simplified axisymmetric representation of the phenomenon. The timeline described above is visualised in Figure~\ref{fig:bos-sequence}, which shows sequential images from the experiment. A movie showing the evolution of the phenomenon is provided in the ancillary files (\verb|fig2_movie.avi|). Details of the experimental methods are provided in the supplementary material.

The resemblance to the well-known rheological phenomenon ``beads-on-a-string'' is clear, and seems to represent an inverse of this case given that the surface stress balance (of air-viscoelastic liquid, including surface tension) in both cases is, in essence, the same. After a period of time the large bubbles on the string collapse, with their point/depth of collapse dependent on the bubble size. The collapse always perturbs the submerged jet flow to some extent (see last frame of Fig \ref{fig:bos-sequence}), but a bubble collapse event which takes place nearer the liquid bath surface can be a significant enough event to entirely disrupt the flow and entry of the plunging jet - effectively halting the submerged flow and coiling process temporarily. The phenomenon can reestablish itself once a stable plunging jet reforms in the settled liquid bath. Moreover, when multiple bubbles-on-a-string are present on the submerged jet, bubbles that come into contact with each other due to differences in their rise velocities can merge together. A rapid increase in the rise velocity is usually observed once they merge, likely owing to the sudden increase in buoyancy. 

The bubbles reduce in size as they rise (as visible in Figure~\ref{fig:bos-sequence}), as air is drawn downwards from the bubble by the falling (submerged) jet. Consequently, and also visible in Figure~\ref{fig:bos-sequence}, the bubble rise velocity reduces with time, as the buoyant forces decrease with decreasing bubble size. This same behaviour is given by a highly simplified model, detailed in the supplementary material, based on mass and momentum balances in a control volume moving with the rising bubble. Key to the rate at which the bubble drains (downwards), and hence the rate at which the bubble rise slows, is the difference in the cross-sectional area of the annular air layer above and below the bubble. In our experiments, the bubbles form at the base of the submerged jet where a coiling instability introduces an asymmetry; we postulate it is this asymmetry, which remains to some degree as the bubble rises, which drives variation in air layer thickness and leads to the bubble draining.

\begin{figure*}%[ht]
\centering
\includegraphics[width=0.99\linewidth]{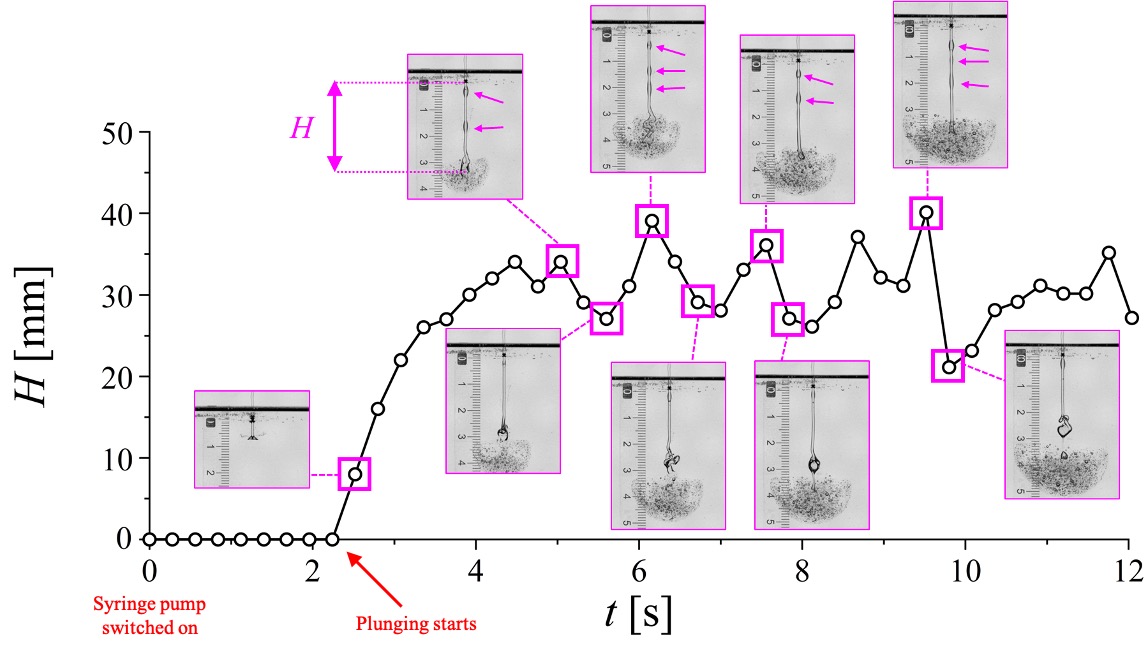}
\caption{Depth reached by jet, $H$, varying with time as for a typical experimental run. The pump is turned on at $t=0$s and liquid entry/plunging occurs at approximately 2.3s.}
\label{fig:plunging dynamics}
\end{figure*}

%\subsection{Plunging Dynamics}\label{plunging}
Figure \ref{fig:plunging dynamics} shows the variation of the depth reached by the submerged jet, $H$, with time, as it evolves over the course of a typical experiment. The inset images and arrows highlight key locations where bubbles-on-a-string form along stable jets of increased length, before bubble collapse disrupts the jet flow, effectively breaking the jet and decreasing effective penetration depth. There is a clear correlation between the number of bubbles and the depth of the jet - the bubbles-on-a-string phenomenon seems to allow the jet to plunge further. There is also a clear periodicity to this behaviour, as the bubbles-on-a-string phenomenon can reestablish itself a short time after a bubble collapse event, with the new stable jet then re-burrowing deeper into the fluid once new bubbles-on-a-string form.

\begin{figure}
         \includegraphics[width=0.32\linewidth]{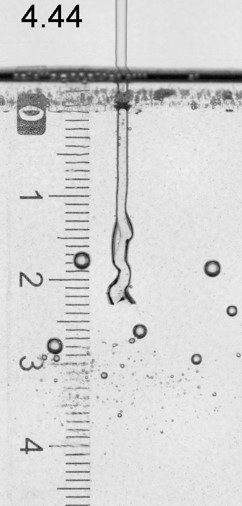}
         \includegraphics[width=0.32\linewidth]{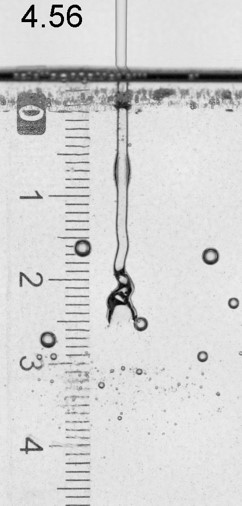}
         \includegraphics[width=0.32\linewidth]{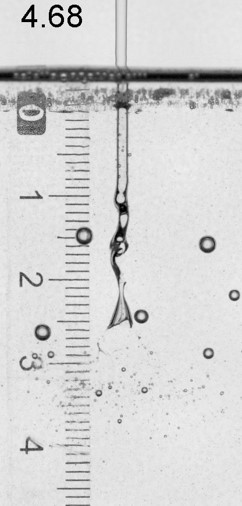}
        \caption{Snapshots of ``bubbles-on-a-string'' life-cycle in a Newtonian surfactant (Glycerol/CAPB) solution: (left) bubble formation (centre) bubble rise (right) bubble collapse. The numbers in the top left of each frame indicate the time in seconds. The depth scale has units of cm.}
        \label{fig:surfactant bos seq}
\end{figure}

\begin{figure}%[ht]
\centering
\includegraphics[width=0.99\linewidth]{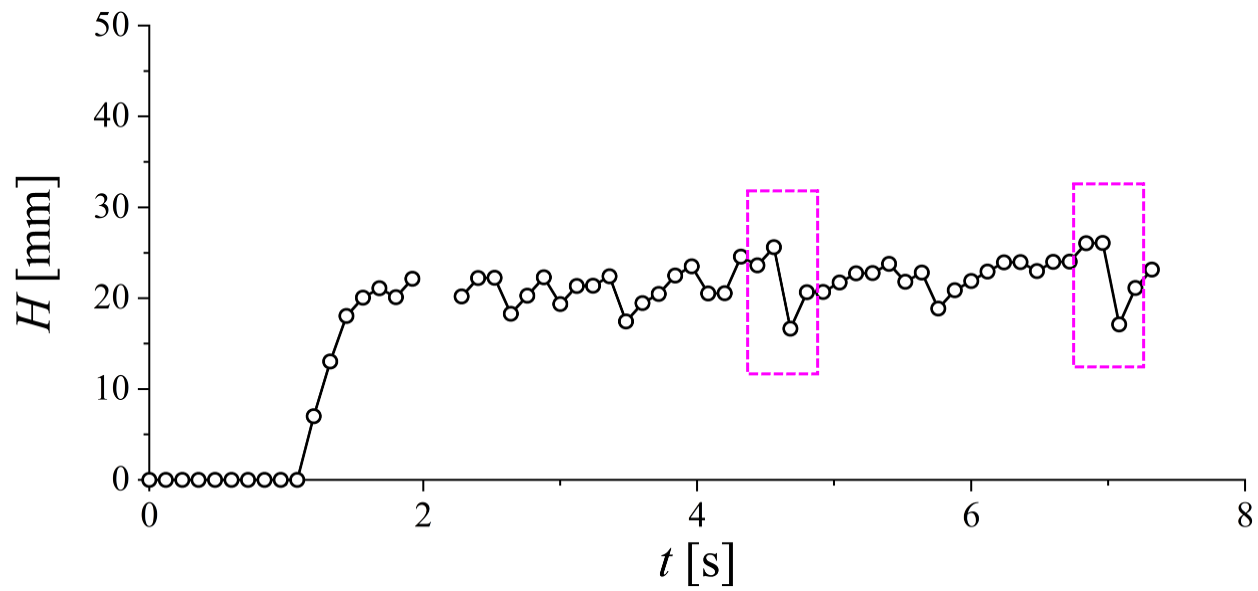}
\caption{Depth reached by jet, $H$, varying with time as for a typical experimental run for the Newtonian surfactant solution. Two separate bubbles-on-a-string life-cycle events are indicated by the dashed boxes.}
\label{fig:jet depth w time for Newt-surfact}
\end{figure}

%\subsection{Newtonian Surfactant Solution}\label{newt}
Fluid surface stresses are key to understanding this phenomenon. To our knowledge, this phenomenon has not been previously observed in the equivalent standard Newtonian flow case. The primary test liquid here is shampoo which, while being viscoelastic, is also a highly concentrated surfactant solution which typically consists of a micellar microstructure, with its viscoelastic and surfactant nature intrinsically coupled. To isolate the role of surfactant effects in this phenomenon, a Newtonian surfactant solution is created and the experiments repeated (note that the distance between the needle and pool was reduced to 34~mm). The Newtonian solution consists of 90\% wt/wt Glycerol, 0.83\% wt/wt CAPB (surfactant), and 9.17\% wt/wt water, formulated to match the surface tension coefficient of the shampoo as closely as possible. Importantly, the bubbles-on-a-string phenomenon is also observed with Newtonian surfactant solution (see Figure \ref{fig:surfactant bos seq}, and ancillary file \verb|fig4_movie.avi|). This suggests the formation of BUoaS relies on low surface tension, with the general life-cycle of the Newtonian surfactant BUoaS being the same as for the shampoo. The plunging dynamics are quite different however, with the Newtonian jet seeming more stable than the shampoo (see Figure \ref{fig:jet depth w time for Newt-surfact}), exhibiting limited submerged coiling (and at reduced amplitude) resulting in fewer (and apparently smaller) BUoaS events. The plunge depth of the Newtonian jet is also lower – this may be a consequence of the shear-thinning/banding nature of the shampoo as it descends and coils through the liquid bath.

%\section{Conclusions}
%\label{sec:conc}
In this work, we have investigated plunging laminar jets into a quiescent liquid bath using a non-Newtonian liquid (shampoo). We have identified a novel phenomenon – ``Bubbles-on-a-string'' (BUoaS) - which to the best of the authors' knowledge has not been reported previously. The bubbles form from coiling activity at the base of the submerged jet, before rising as toroidal bubbles around the jet axis. The bubbles have complex dynamics (involving merging and collapsing) which strongly affect the jet plunging depth and hence air entrainment depth. We also observe a subdued version of BUoaS with a Newtonian surfactant solution (Glycerol/CAPB), suggesting that low surface tension is key to the phenomenon, but viscoelasticity and/or shear-thinning appear to create the optimal conditions for a pronounced and persistent phenomenon, in part by affecting plunging dynamics and increasing jet penetration depth. A comprehensive understanding of the BUoaS phenomenon will have implications across manufacturing and processing industries, where air entrainment leads to a reduction or improvement in product quality. Further work on this topic is ongoing and will involve detailed characterisation of the conditions and parameter regimes (based on Reynolds, Weber, and Weissenberg numbers) needed for the formation of BUaoS. In this regard, it will be particularly important to assess the role of shear-thinning effects and elasticity on BUoaS and plunging dynamics. 

\section*{Acknowledgments}

We acknowledge funding granted from the University of Manchester’s Department of Chemical Engineering for this work. JK acknowledges funding from the Royal Society via a University Research Fellowship (URF\textbackslash R1\textbackslash 221290). We would like to thank the University of Manchester MECD workshop for help with the experiments. 

\bibliography{sample}
\end{document}